# The Botization of Science? Large-scale study of the presence and impact of Twitter bots in science dissemination

Wenceslao Arroyo-Machado[1], Enrique Herrera-Viedma[2] & Daniel Torres-Salinas[1]

wences@ugr.es; viedma@ugr.es; torressalinas@ugr.es

[1]Department of Information and Communication, University of Granada

[2]Department of Computer Science and Artificial Intelligence, University of Granada

**Abstract**

Twitter bots are a controversial element of the platform, and their negative impact is well known. In the field of scientific communication, they have been perceived in a more positive light, and the accounts that serve as feeds alerting about scientific publications are quite common. However, despite being aware of the presence of bots in the dissemination of science, no large-scale estimations have been made nor has it been evaluated if they can truly interfere with altmetrics. Analyzing a dataset of 3,744,231 papers published between 2017 and 2021 and their associated 51,230,936 Twitter mentions, our goal was to determine the volume of publications mentioned by bots and whether they skew altmetrics indicators. Using the BotometerLite API, we categorized Twitter accounts based on their likelihood of being bots. The results showed that 11,073 accounts (0.23% of total users) exhibited automated behavior, contributing to 4.72% of all mentions. A significant bias was observed in the activity of bots. Their presence was particularly pronounced in disciplines such as Mathematics, Physics, and Space Sciences, with some specialties even exceeding 70% of the tweets. However, these are extreme cases, and the impact of this activity on altmetrics varies by speciality, with minimal influence in Arts & Humanities and Social Sciences. This research emphasizes the importance of distinguishing between specialties and disciplines when using Twitter as an altmetric.

**Highlights**

- Bots account for 0.23% of Twitter users, yet they produce 11.53% of tweets mentioning papers.
- Bots have a greater presence in the hard sciences, with Mathematics being the most affected discipline with 39.4% of the tweets being from them.
- The tweets from bots to papers are not always distributed evenly, and in some specialties, they influence the altmetric impact.
- In Arts & Humanities and Social Sciences, the impact of bots on altmetrics is minimal.

## 1. Introduction

Bots, machines designed to mimic human behavior, are far from being a novel idea (Turing, 1950). What is novel is their materialization and application in real-world scenarios. With advancements in computing, it did not take long for these algorithms to emerge in different realms of the digital world, giving rise to an entire ecosystem of bots (Leonard, 1997). There are bot applications with varied impacts and functionalities, ranging from chatbots that interact with users through natural language conversations (Følstad & Brandtzæg, 2017) further popularized with the advent of ChatGPT (Vallance, 2022), to spambots that massively distribute unwanted content (Ferrara, 2018). Moreover, bot applications on social media are no exception (Ferrara et al., 2016), where the so-called social media bots have become common tools, performing varied tasks such as content generation and automated interactions with other human users or bots (Gorwa & Guilbeault, 2020). Their presence and implications are also varied in this specific context; although they can be found on many platforms such as Wikipedia (Zheng et al., 2019), it is on Twitter where they have prominently manifested, also gaining sophistication over time (Alothali et al., 2018).

The presence of social bots on Twitter is a contentious issue. Quantifying them poses multiple and interrelated challenges. One of the challenges arises from the difficulty of establishing a common sharing profile or one that considers the variety of behaviors of bots (Rodríguez-Ruiz et al., 2020). This gives rise to another challenge, which is the equally varied range of proposals available for their estimation (Fukuda et al., 2022; Tan et al., 2023). It is due to these issues that discrepancies have arisen in the provided figures; for instance, Twitter itself estimates that the prevalence of bots is around 8% (Subrahmanian et al., 2016), while the current owner, Elon Musk, estimated 33% prior to his purchase of the platform (Conger, 2022). Other estimates suggest a value close to 15%, a midpoint between Twitter's and Musk's estimates (Varol et al., 2017). Delving into the complex ecosystem of Twitter bots, two types of schemes become apparent for classifying this variety of bots and behaviors: one based on the intentions behind their creation (Ferrara et al., 2016) and another on the degree to which they mimic human behavior (Stieglitz et al., 2017). The first scheme is noteworthy as it establishes a fundamental distinction between benign and malicious bots, with the latter generating more interest in research due to their potential consequences (Orabi et al., 2020).

Addressing this group of malicious bots, we can find notorious cases in different areas. The political arena is one of the most affected. The role of bots in influencing electoral outcomes has been widely discussed, evidencing their activity in multiple electoral processes around the world (Bovet & Makse, 2019; Gorodnichenko et al., 2021; Pastor-Galindo et al., 2020; Santana & Huerta Cánepa, 2019). In relation to this, the spread of fake news is another hot spot of malicious bot activity, especially for propagandistic purposes (Al-Rawi et al., 2019; Jones, 2019; Shao et al., 2018). Within this context of destabilization and alteration of public opinion, the risks they pose to public health have also been analyzed, such as recommending harmful products or encouraging anti-vaccine debate (Allem et al., 2017; Broniatowski et al., 2018). Similarly, Twitter bots have been identified behind troll behaviors (Alsmadi & O'Brien, 2020), cybersecurity crimes (Shafahi et al., 2016), and even stock market manipulation (Fan et al., 2020). However, despite the concern raised by these automated accounts, it is fair and necessary to disassociate the negative connotations from the term bot as there are also multiple examples of benign activity, and their behavior differs from the malicious ones.

A field where Twitter bots have found practical and beneficial applications is in science and scientific communication. For instance, in discussions surrounding vaccines in the context of COVID-19, there has been active participation of bots conveying a positive perception about their efficacy (M. Zhang et al., 2022). They have even proven to be beneficial in amplifying the visibility of scholarly outputs by tweeting them (Ye & Na, 2020). Indeed, the presence of bots alerting about preprint submissions to repositories like arXiv is well-known (Shuai et al., 2012). This activity, where bots tweet mentioning a scholarly output, is not inherently malicious, and many of these accounts openly identify as bots, and their scripts are public (Kollanyi, 2016). Yet, this does not imply that the activities of these accounts are without consequences. One of the classical challenges of using altmetrics to evaluate social attention is their susceptibility to manipulation, thereby skewing the perception of said attention (Thelwall, 2014). This is a risk that is not unfounded and, in fact, has been warned against upon finding bots among the main accounts that tweet publications from areas such as Dentistry (Robinson-Garcia et al., 2017) or Microbiology (Robinson-Garcia et al., 2019), and even video publications (Xu et al., 2018). Nonetheless, there are very few studies focused on the detection and analysis of such bots, with only preliminary approaches having been undertaken that warn of the risk these accounts can pose to altmetrics by easily boosting the number of tweets (Haustein et al., 2016). These concerns have been taken into consideration in recent methodological proposals to distinguish the attention of bots from other Twitter audiences (Arroyo-Machado & Torres-Salinas, 2023). However, the global presence of Twitter bots in research dissemination is still unknown, as well as whether they pose a risk to altmetrics.

In this paper, we aim to explore the impact of Twitter bot accounts that automatically mention publications, and to that end, we have formulated the following research questions:
- What volume of publications are mentioned by bots on Twitter?
- Do Twitter bots alter altmetrics indicators by artificially and unevenly increasing mentions of publications?

To address these research questions, our main objective is to provide the first large-scale quantitative estimate of the presence of Twitter bots in the context of scientific communication and to determine their level of influence on altmetrics. The specific objectives set for this study are as follows:
- Objective 1. To determine the presence of bots and their activity in the dissemination of papers on Twitter globally, differentiating by scientific disciplines and specialties.
- Objective 2. To study the impact of Twitter bots on the altmetrics of publications and identify the specialties and disciplines they influence the most.

## 2. Methodology

### 2.1. Data

The collection of bibliographic records and their mentions on Twitter was conducted on September 5, 2022. All papers published between 2017 and 2021 indexed in the Science Citation Index (SCIE), Social Science Citation Index (SSCI), and Art & Humanities Citation Index (AHCI) were retrieved from Web of Science, totaling 9,141,397 papers. For each of these publications, we obtained their Web of Science categories, a scheme encompassing 254 scientific specialties. Furthermore, through the mapping scheme proposed by Arroyo-Machado & Torres-Salinas (2021), we obtained the ESI field of the papers from the Web of Science categories, a schema that includes 22 major disciplines plus the Multidisciplinary category.

**Table 1**. Distribution of Web of Science papers published between 2017 and 2021 and mentions on Twitter by ESI field

| ESI field | Papers | Web of Science categories | Tweeted papers | % | Tweets | Tweeters |
|---|---|---|---|---|---|---|
| *Agricultural Sciences* | 365,928 | 9 | 121,913 | 33% | 976,556 | 264,669 |
| *Arts & Humanities* | 222,277 | 24 | 61,725 | 28% | 636,914 | 205,128 |
| *Biology & Biochemistry* | 483,343 | 7 | 278,533 | 58% | 4,178,047 | 779,387 |
| *Chemistry* | 1,208,833 | 10 | 401,034 | 33% | 1,771,965 | 258,478 |
| *Clinical Medicine* | 1,901,366 | 44 | 1,051,725 | 55% | 18,374,693 | 2,472,525 |
| *Computer Science* | 486,363 | 9 | 92,399 | 19% | 676,381 | 191,562 |
| *Economics & Business* | 218,860 | 5 | 86,408 | 39% | 1,081,441 | 326,141 |
| *Engineering* | 1,803,652 | 28 | 316,740 | 18% | 1,708,945 | 487,948 |
| *Environment/Ecology* | 587,529 | 5 | 258,850 | 44% | 3,376,896 | 620,641 |
| *Geosciences* | 442,595 | 11 | 161,684 | 37% | 1,606,015 | 348,299 |
| *Immunology* | 177,426 | 3 | 120,164 | 68% | 2,403,361 | 649,596 |
| *Materials Science* | 940,022 | 10 | 217,656 | 23% | 836,899 | 180,358 |
| *Mathematics* | 344,838 | 5 | 48,776 | 14% | 261,983 | 72,812 |
| *Microbiology* | 289,595 | 4 | 176,707 | 61% | 2,280,502 | 518,192 |
| *Molecular Biology & Genetics* | 260,633 | 4 | 166,921 | 64% | 3,216,259 | 634,625 |
| *Multidisciplinary* | 334,722 | 1 | 224,568 | 67% | 8,103,923 | 1,791,621 |
| *Neuroscience & Behavior* | 304,535 | 4 | 206,580 | 68% | 2,571,154 | 509,091 |
| *Pharmacology & Toxicology* | 288,200 | 3 | 135,439 | 47% | 1,052,946 | 303,864 |
| *Physics* | 904,321 | 11 | 312,992 | 35% | 1,295,154 | 212,204 |
| *Plant & Animal Science* | 456,182 | 11 | 234,974 | 52% | 2,436,689 | 363,144 |
| *Psychiatry/Psychology* | 331,955 | 14 | 214,347 | 65% | 3,064,398 | 712,607 |
| *Social Sciences, General* | 485,079 | 31 | 265,999 | 55% | 3,845,651 | 802,363 |
| *Space Sciences* | 109,748 | 1 | 78,455 | 71% | 512,318 | 91,355 |
| **Total** | 9,141,397 | 254 | 3,744,231 | 41% | 51,230,936 | 4,872,316 |

From the data from Web of Science, we used the DOIs of the papers to query Altmetric.com and retrieve all Twitter mentions, encompassing both tweets and retweets. This process was also conducted on September 5, 2022, retrieving all Twitter mentions for all papers up to that date. As a result, a total of 51,230,936 Twitter mentions were obtained, made by 4,872,316 users for 3,744,231 papers (Table 1). From this dataset, the analysis was conducted.

## 2.2. Methods

In this study, our aim was to identify Twitter accounts that exhibit clear and overt automated behavior, such as those functioning as paper repository feeds or those that explicitly state in their description that they engage in automated activity. We did not seek to identify other types of social bots, like those mimicking human behavior, whose identification is more ambiguous and challenging. Consequently, we utilized the BotometerLite API (Yang et al., 2020), a machine learning tool that employs a more streamlined model than the primary Botometer API V4 (Sayyadiharikandeh et al., 2020). This tool focuses on predicting large data volumes but simplifies the features utilized for account classification, considering basic elements like tweet count, number of friends, and whether the profile has been personalized. Additionally, it evaluates derivatives of these attributes, such as tweet frequency and the follower-friend ratio, ultimately achieving efficient bot identification at a reduced computational expense (Yang et al., 2020).

On December 26, the BotometerLite API was queried using identifiers of Twitter accounts that mention papers. For each of the 4,872,316 accounts, a "botscore" was retrieved, which indicates the extent to which an account displays human or bot behavior. The botscore ranges from 0 to 1, with 0 indicating human-like behavior and 1 indicating bot-like behavior. A primary challenge with the botscore is setting a threshold to determine which accounts are bots (Yang et al., 2022). Various suggestions exist for this, though they all relate to the main Botometer model. Given this scenario, various tests have been conducted to select the most suitable botscore threshold. This metric is paired with other variables that help correct Botometer's shortcomings, such as false positives in older accounts or those currently inactive (Rauchfleisch & Kaiser, 2020). In addition, these adjustments are intended to improve their performance when applied to a specific context such as science dissemination. In this manner, a robustness check was performed (Appendix A1) by extracting a representative sample of our dataset and using other datasets with the goal of conducting different tests and selecting the optimal criteria for detecting bots that mention publications while reducing the risk of false positives. As a result, it was decided to use as a criterion to label bots those tweeters who have a botscore above 0.6 and who made more than 8 mentions.

After delineating the bots, a descriptive study of their presence and impact on the dissemination of science on Twitter was conducted using Python and R. Firstly, using descriptive statistics, we provided a general overview of the presence of these automated accounts and the volume of mentions, tweets, and retweets attributed to them, differentiating by both discipline (ESI field) and specialty (Web of Science category). For this, we used as indicators the percentage of Twitter accounts that are bots and the percentages of mentions and tweets made by bots. Secondly, the impact of bots was determined. Here, the impact is understood as the bot interference on social attention-measured through Twitter activity (Thelwall, 2020a), where in the worst scenario it could inflate the attention to papers that otherwise would not receive it, or in the best case distribute it evenly without significantly altering such attention. In this case, the analysis was conducted at the paper level, aggregating by discipline and specialty, and focusing exclusively on tweets by removing the social component of Twitter, that is, the retweets. To delimit the impact, various metrics were employed to gain a clear picture of this phenomenon. Through the average percentage of bot tweets and the GINI index, it was possible to discern disciplines and specialties with a high presence of automated activity and an uneven distribution. Analyzing the differences between bot tweets and those of other twitters (no-bots) allowed for the identification of the impact. By examining the average percentile difference of both values and the Spearman correlation between them, we could identify the overall bot impact and those areas with a more pronounced effect. Specifically, disciplines and specialties with a high average percentile difference and a low correlation coefficient.

The list of bots identified in this analysis is openly shared through Zenodo (doi:10.5281/zenodo.8075234) while the Python and R scripts with all the data processing and analysis can be found on GitHub (https://github.com/Wences91/botization_science).

## 3. Results

### 3.1. Presence of Twitter bots in Scientific Communication

In total, there are 11,073 accounts with automated behavior, representing 0.23% of all Twitter users analyzed. Although bots constitute a minority of users engaging in the global discussion on science, their activity is noticeably present. They carry out a total of 2,420,650 mentions (4.72% of the total), with their presence mainly reflected in the tweets, for which they are responsible for 2,016,607 (11.53% of the total), compared to 402,043 retweets (1.19% of the total). Bots mention an average of 182.51 papers ($\pm1063.38$) and make an average of 218.61 mentions ($\pm1279.11$), 182.3 tweets ($\pm1212.47$), and 36.31 retweets ($\pm324.48$). These figures reveal a significant bias in the bots' activity, something that is noticeable when observing the bots with the most mentions (Table S3), as the top 20 accounts for 19% of all bot-made mentions and 21% of the tweets. In this context, the accounts Behav Ecol Papers (43,538 mentions), Symbiosis papers (32,792), and POPapers (31,088) stand out.

However, presence and involvement vary according to the speciality (Figure 1). By looking at the classification provided by the ESI fields, such differences become apparent. It is in specialties related to hard sciences, especially Mathematics, Physics, and Space Sciences, where the presence of bots not only exceeds the values of other ones but also significantly stands out in the percentage of their activity over the Twitter discussion. In Mathematics, bot presence is most striking, as 1.7% of the Twitter users who mention papers in the field are bots, and they are responsible for 22.1% of the mentions and 39.4% of the tweets. These numbers contrast sharply with those of Arts & Humanities and Social Sciences, General, where bots represent 0.6% and 0.4% respectively, making up 3% of the tweets in both cases. Regarding the presence of bots in each ESI field, it is worth noting that in all cases, these values are above the global presence, as 91% of the bots mention papers from more than one ESI field, with an average of 7 ESI fields mentioned by the bots.

**Figure 1**. Percentages of Twitter users mentioning publications that are bots, and percentages of mentions and tweets made by bots, differentiated by ESI fields

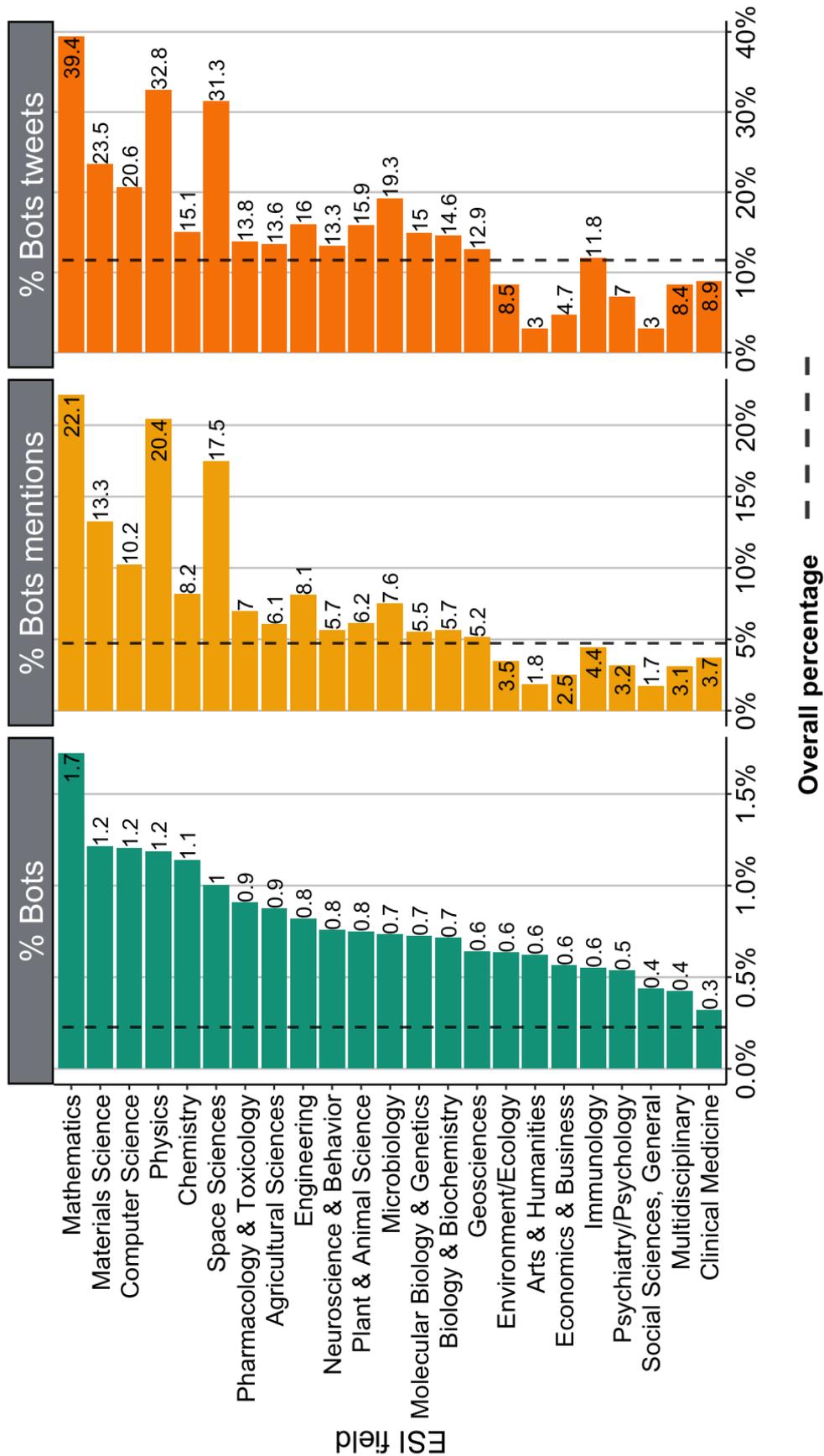

By delving into the fields and observing the presence of bots and their activity by Web of Science category, it is possible to identify the specialties in which they are most prevalent (Figure 2). In this case, Mathematics, Mathematics, Applied, and Physics, Mathematical stand out with 71.69%, 64.07%, and 58.14% of tweets being made by bots respectively. This means that bot activity in these categories constitutes the majority, although bots only make up 4%, 3.57%, and 3.13% of Twitter users. Bots represent a limited presence in all categories, although their activity varies among them. Microscopy, with 8.96% bot users, has the highest presence of automated users, followed by Logic (6.63%), Materials Science, Ceramics (5.46%), Materials Science, Characterization & Testing (5.21%), and Literature, Slavic (5.21%). The last category is remarkable because, despite having a noticeable presence of bots, their mentions only account for 3.26% and their tweets for 0.51%. However, this category only has 368 mentions to 103 papers. In the case of broad Arts and Humanities categories, both presence and activity are limited. For example, in History, 366 out of 61,965 Twitter users are bots (0.59%) and 875 out of 53,391 tweets (1.64%) are made by them.

**Figure 2**. Distribution of Web of Science categories according to the total tweets mentioning their papers and the percentage of these that are made by bots

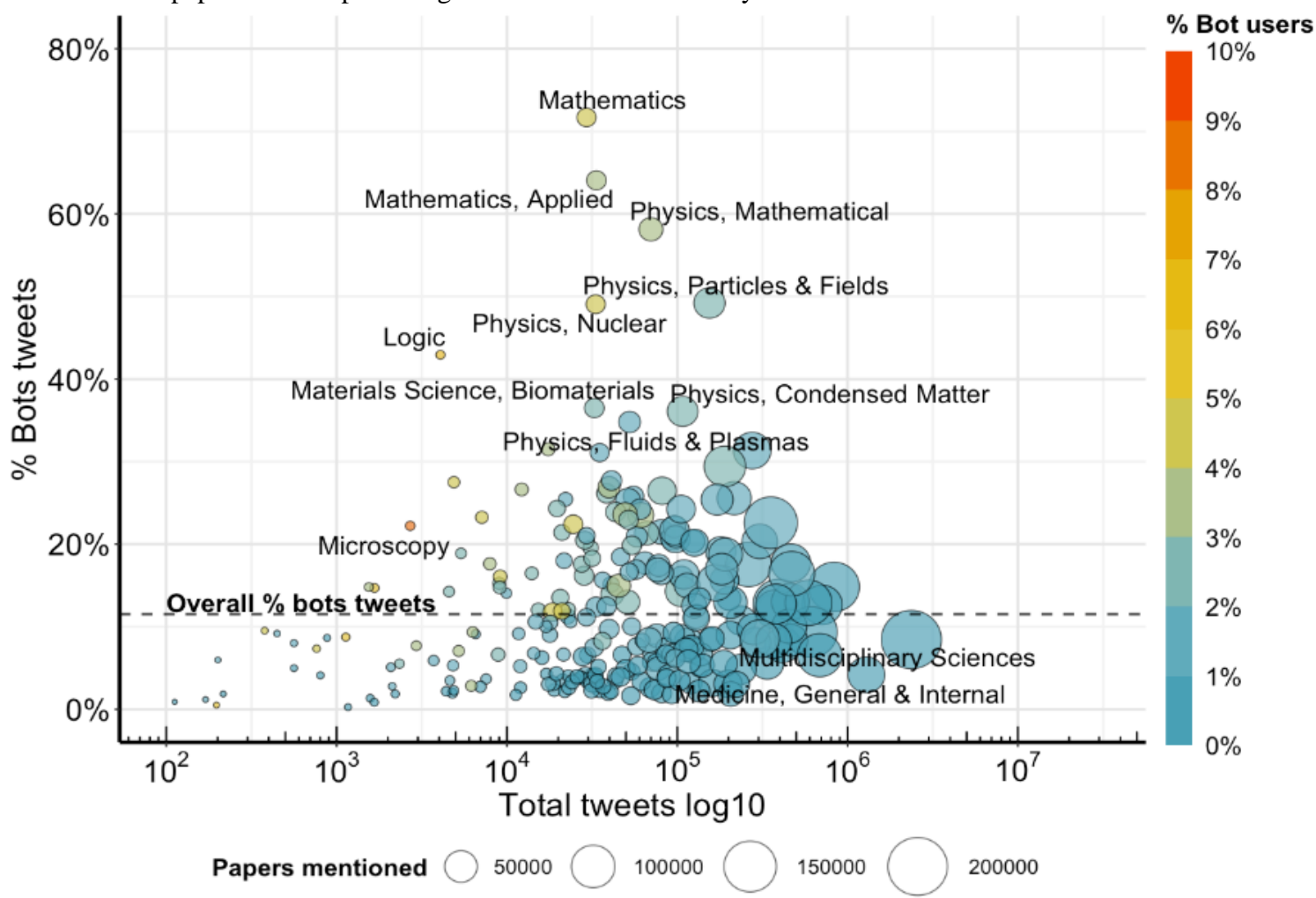

## 3.2. Impact of Twitter bots on altmetrics

Despite the presence that bots might have in the total mentions or tweets received by papers in a general ESI field or a specific Web of Science category, the impact can vary. Here, the impact is understood as the effect on altmetrics indicators, in particular the tweet count. In the worst-case scenario, it could inflate the tweet count for publications that would otherwise not receive it, and in the best-case scenario, distribute it evenly without significantly altering the order. In this way, after distinguishing in each paper the tweets mentioning them, published by bots and non-bots, we were able to define and specify the impact of bots, highlighting research areas with a notable and uneven artificial inflation of tweets that could lead to a manipulation of altmetrics.

**Table 2**. Percentages and distribution of bot tweets and relationship metrics between bot and non-bot tweets at the paper level, across ESI fields

| In this table the metrics are calculated at paper level | *Bot tweets percentage* | | *Percentage of papers with bot tweets* | | | *Relationship between bot and non-bot tweets* | |
|---|---|---|---|---|---|---|---|
| | Average | Gini index | > 0% | >= 50% | == 100% | Average percentile diff. | Spearman correlation↓ |
| *Overall* | 20.25% | 0.77 | 34% | 21% | 13% | 10.55 | 0.868 |
| *Mathematics* | 44.56% | 0.55 | 54% | 46% | 37% | 23.41 | 0.419 |
| *Physics* | 36.63% | 0.60 | 51% | 41% | 25% | 18.61 | 0.641 |
| *Materials Science* | 28.93% | 0.70 | 37% | 32% | 23% | 16.69 | 0.700 |
| *Engineering* | 20.50% | 0.79 | 27% | 22% | 16% | 13.51 | 0.759 |
| *Space Sciences* | 41.11% | 0.52 | 67% | 46% | 22% | 14.05 | 0.768 |
| *Computer Science* | 19.91% | 0.77 | 31% | 21% | 12% | 13.00 | 0.780 |
| *Pharmacology & Toxicology* | 25.10% | 0.73 | 36% | 28% | 17% | 13.94 | 0.791 |
| *Chemistry* | 17.45% | 0.81 | 25% | 19% | 12% | 11.77 | 0.819 |
| *Agricultural Sciences* | 20.35% | 0.78 | 30% | 22% | 15% | 11.91 | 0.824 |
| *Microbiology* | 32.18% | 0.63 | 54% | 34% | 19% | 12.53 | 0.831 |
| *Geosciences* | 17.21% | 0.80 | 29% | 19% | 10% | 10.95 | 0.846 |
| *Immunology* | 27.46% | 0.68 | 47% | 29% | 16% | 11.57 | 0.853 |
| *Biology & Biochemistry* | 25.96% | 0.70 | 46% | 27% | 16% | 11.09 | 0.867 |
| *Plant & Animal Science* | 22.78% | 0.74 | 39% | 24% | 14% | 10.93 | 0.870 |
| *Molecular Biology & Genetics* | 28.57% | 0.66 | 53% | 29% | 16% | 10.18 | 0.884 |
| *Neuroscience & Behavior* | 20.12% | 0.75 | 39% | 20% | 11% | 9.34 | 0.903 |
| *Clinical Medicine* | 17.86% | 0.80 | 31% | 19% | 11% | 8.76 | 0.904 |
| *Environment/Ecology* | 13.16% | 0.84 | 24% | 14% | 7% | 7.64 | 0.916 |
| *Multidisciplinary* | 20.83% | 0.72 | 48% | 20% | 10% | 7.82 | 0.928 |
| *Psychiatry/Psychology* | 10.94% | 0.86 | 24% | 11% | 5% | 6.02 | 0.946 |
| *Economics & Business* | 6.65% | 0.92 | 12% | 7% | 4% | 4.07 | 0.958 |
| *Arts & Humanities* | 4.24% | 0.95 | 8% | 4% | 3% | 2.80 | 0.970 |
| *Social Sciences, General* | 4.27% | 0.94 | 10% | 4% | 2% | 2.56 | 0.979 |

At the ESI field level, we again find an unequal distribution in terms of the average percentage of bot tweets that papers receive (Table 2). The hard science fields stand out once again, with Mathematics (44% average bot tweets) and Space Sciences (41%) having the highest presence, double the general average (20%). Through the GINI index of the percentage of bot tweets, areas with a higher concentration of this type of tweets can be identified. Such is the case for Arts & Humanities and General Social Sciences, which have a very low average of bot tweets (in both cases the average percentage of bot tweets is 4%) and are concentrated in a minority of works (the GINI index is 0.95 and 0.94 respectively). Space Sciences is the field that reflects a more equitable distribution with a GINI index of 0.52, followed by Mathematics at 0.55. This highlights the high presence of bot tweets in these fields. In both cases, 46% of the papers owe at least half of their tweets to bots, something that might explain these results.

However, this does not necessarily imply a high impact but rather highlights the presence and distribution of bot tweets. It is when directly comparing the bot and non-bot tweets of the papers that this impact becomes more apparent. Firstly, the average difference in percentiles between bot and non-bot tweets makes it possible to pinpoint ESI fields where bots have a considerable impact, such as Mathematics (with an average percentile difference of 23.41), Physics (18.61), and Materials Science (16.69). The general average percentile difference is 10.55. Secondly, the Spearman correlation coefficient between bot and non-bot tweets underscores this impact. The lowest correlations again point to Mathematics ($\rho$=0.419), Physics ($\rho$=0.641), and Materials Science ($\rho$=0.700), indicating a weak relationship between bot and non-bot tweets. That is, papers tweeted by bots are not tweeted by non-bots, and vice versa. The rest of the ESI fields are above $\rho$=0.750, while the general level is $\rho$=0.868. Hence, we can identify a reduced impact at a general level, but it varies depending on the field, especially affecting Mathematics and to a lesser extent some science fields.

The study of the impact at the Web of Science category level highlights specific specialties most affected by bots (Table 3). Both Technology and especially Physical Sciences categories encompass specialties significantly influenced by bots. Notably, Mathematics, Applied ($\rho$=0.023) and Mathematics ($\rho$=0.041) show almost a complete lack of correlation between bot and non-bot tweets, with average percentile variations around 30 and more than half of the papers being tweeted only by bots. In Life Sciences & Biomedicine, several specialties exhibit a moderate bot impact, with Agricultural Engineering ($\rho$=0.595) standing out. For Social Sciences, no specific category indicates bot impact, this can be seen in Psychology, Mathematical ($\rho$=0.860), its most affected speciality. In Arts & Humanities, the scenario is the same, with Literature, German, Dutch, Scandinavian ($\rho$=0.877), and Literary Theory & Criticism ($\rho$=0.886) being the most impacted.

**Table 3**. Top 5 Web of Science categories by main area most affected by bot tweets

| Main area | Bot tweets percentage | Percentage of papers with bot tweets | | Relationship between bot and non-bot tweets | |
|---|---|---|---|---|---|
| ***Arts & Humanities*** | Average | ><br>0% | ==<br>100% | Average percentile diff. | Spearman correlation↑ |
| *Literature, German, Dutch, Scandinavian* | 10.95% | 14% | 9% | 7.65 | 0.877 |
| *Literary Theory & Criticism* | 10.18% | 11% | 9% | 6.94 | 0.886 |
| *Literary Reviews* | 15.40% | 20% | 12% | 8.91 | 0.891 |
| *Religion* | 4.71% | 7% | 3% | 4.60 | 0.914 |
| *Architecture* | 5.98% | 8% | 5% | 4.74 | 0.919 |
| ***Life Sciences & Biomedicine*** | Average | ><br>0% | ==<br>100% | Average percentile diff. | Spearman correlation↑ |
| *Agricultural Engineering* | 29.37% | 34% | 26% | 18.58 | 0.595 |
| *Food Science & Technology* | 25.94% | 34% | 20% | 16.18 | 0.704 |
| *Dentistry, Oral Surgery & Medicine* | 25.59% | 34% | 20% | 15.75 | 0.715 |
| *Agriculture, Dairy & Animal Science* | 32.89% | 44% | 25% | 16.16 | 0.726 |
| *Otorhinolaryngology* | 35.35% | 47% | 27% | 15.73 | 0.739 |
| ***Physical Sciences*** | Average | ><br>0% | ==<br>100% | Average percentile diff. | Spearman correlation↑ |
| *Mathematics, Applied* | 58.92% | 64% | 53% | 30.27 | 0.023 |
| *Mathematics* | 70.91% | 75% | 65% | 29.50 | 0.041 |
| *Physics, Mathematical* | 63.81% | 80% | 47% | 20.71 | 0.482 |
| *Physics, Fluids & Plasmas* | 45.23% | 57% | 35% | 19.97 | 0.569 |
| *Chemistry, Applied* | 32.16% | 38% | 27% | 18.54 | 0.627 |
| ***Social Sciences*** | Average | ><br>0% | ==<br>100% | Average percentile diff. | Spearman correlation↑ |
| *Psychology, Mathematical* | 25.83% | 47% | 14% | 11.65 | 0.860 |
| *Psychology, Biological* | 29.96% | 55% | 18% | 9.13 | 0.905 |
| *Psychology, Developmental* | 12.59% | 27% | 6% | 7.23 | 0.928 |
| *Psychology, Psychoanalysis* | 6.83% | 10% | 5% | 5.08 | 0.934 |
| *Education, Special* | 10.50% | 20% | 6% | 6.03 | 0.944 |
| ***Technology*** | Average | ><br>0% | ==<br>100% | Average percentile diff. | Spearman correlation↑ |
| *Mechanics* | 41.94% | 49% | 36% | 23.08 | 0.412 |
| *Telecommunications* | 18.40% | 21% | 15% | 16.54 | 0.544 |
| *Automation & Control Systems* | 18.57% | 21% | 15% | 16.25 | 0.559 |
| *Engineering, Electrical & Electronic* | 20.11% | 25% | 15% | 16.76 | 0.581 |
| *Materials Science, Biomaterials* | 39.83% | 48% | 33% | 18.44 | 0.625 |

## 4. Discussion

In this paper, we conduct a large-scale analysis of the presence of bots in the global discussion of science on Twitter. For this, mentions made from Twitter to papers published between 2017 and 2021 and indexed in Web of Science have been studied, which amounts to a total of 51 million mentions to 3.7 million papers. The results show that while these accounts constitute less than 1% of the nearly 5 million accounts analyzed, they are responsible for approximately 12% of the 17 million tweets examined. Moreover, these findings are not consistent across different scientific disciplines and specialties, with Mathematics emerging as a particularly affected area. Regarding the impact that bots have on altmetrics, the results once again display variability depending on the scientific area, with the categories of Arts & Humanities and Social Sciences being the least affected. These findings are expected to be useful for altmetric studies and to contribute to a better characterization of the audience that scientific activity arouses on Twitter.

Considering the uneven presence of mentions and bot tweets across ESI fields, where the hard sciences particularly stand out against Arts & Humanities and Social Sciences, one initial explanation for this phenomenon could be the differences in communication channels of each scientific discipline (Torres-Salinas et al., 2023). Fields such as Mathematics, Physics, Materials Science, or Space Sciences, where a greater presence and impact of bots have been identified, align with research fields that show less activity on this social media. This might suggest that bots are filling a void not heavily exploited by their respective communities. However, Arts & Humanities break this mold as Twitter is also an infrequent channel for their communication. Therefore, another complementary reason for this occurrence might be open access and the presence of bots that serve as alert systems for repositories like arXiv or bioRxiv. While it is true that there is a prevailing trend of bot attention towards papers published under open access, especially in 'green' modalities, this trend is also present in Arts & Humanities, and Social Sciences, General (Table S2). Thus, in the absence of more comprehensive research into the reasons, the findings of this study suggest that multiple factors drive the presence and impact of Twitter bots in research dissemination.

Nevertheless, the presence of the bots analyzed here should not be viewed as a risk or threat for scientific communication. The main bots in terms of paper mentions do not conceal personal or harmful interests. Many of them conduct tracking and monitoring of papers to function as alert systems. In fact, in the case of Mathematics, the Web of Science category with the greatest bot presence, this same trend has been identified. Bots actually represent a valuable opportunity in this instance to disseminate and grant greater visibility to research that, otherwise, would not have, as evidenced by the 65% of Mathematics papers that owe all their tweets to bots. This role of bots as drivers of visibility and discussion of papers has been previously analyzed by Ye et al. (2022), who concluded that, in the case of COVID-19, bots played a significant role in facilitating the dissemination of research on Twitter. While these bots might not inherently be negative, it does not imply that this type of activity should be overlooked, especially in contexts like evaluative altmetrics (Arroyo-Machado & Torres-Salinas, 2023). These results not only highlight but also provide clear evidence that the fundamental risk of manipulation or inflation of metrics by Twitter bots, whether deliberate or not, within altmetrics is a genuine concern.

Can we conclude that using Twitter as an altmetric source should be avoided? Not at all, though this answer requires several clarifications. First and foremost, far from invalidating tweet counts as a measure of social attention, these results support the idea that altmetrics should be used selectively depending on the scientific area (Thelwall, 2020b). For instance, in Art & Humanities and Social Sciences, the impact of Twitter bots appears to be minimal, thus making tweet counting a secure metric in this context. In contrast, its use in Mathematics and Physics would be strongly discouraged due to the notably higher susceptibility to bot influence in these fields. Secondly, bots are just one of the multiple audiences that can be present on Twitter, and eliminating them does not mean focusing exclusively on the broad public audience. The next challenge involves identifying academic tweeters, as they represent one of the main audiences (L. Zhang et al., 2023), and this may determine whether we are discussing scientific communication or social attention (Haustein, 2019). For this reason, these contributions should not be seen as deterrents from using tweets as an altmetric, but as progress in understanding the utility and risks of using this metric to adjust its application to relevant scenarios. Indeed, there is no single perfect and universal metric. The varied metrics for research evaluation, including the push for altmetrics, reflect the need to fully capture how a work is being utilized (Aubert Bonn & Bouter, 2023).

This work is not without limitations. Firstly, it is pertinent to highlight that the results being presented in this study are applicable to Twitter and allow for a better understanding of how bots functioned in this context. However, following the acquisition of the social medium by Elon Musk and its transformation into X, these results may not fully conform to this new reality. This is primarily due to various restrictions that prevent users from developing bots for free. Secondly, the use of BotometerLite firstly involves utilizing a simpler learning model than Botometer and, alongside it, may provide varied responses to minimal changes or inaccuracies in its answers to specific characteristics, such as older accounts. Thirdly, it is also necessary to highlight the problems inherent to altmetrics studies, notably the loss of tweets or the non-capture of those tweets that do not include a publication identifier, such as the DOI or the URL.

## Appendix

### A1. Robustness check

To conduct the tests, we initially drew a random sample adjusted to a 99% confidence level with a 4% margin of error, resulting in a total of 1,034 tweeters. Each of these accounts underwent a manual review between January 20 and 27, 2023, to classify them as bot or not. During this process, 20 accounts that were either blocked or deleted at the time of review were removed. The general criteria set for this identification are detailed in Table 1. However, this initial sample only identified 6 bots, accounting for a mere 0.6%. To achieve a more representative sample, from the total accounts in the dataset, 166 accounts were validated as bots by searching for descriptions containing terms like "bot", "automated", "arxiv", or "papers". Consequently, the dataset used for testing comprises 1180 tweeters, of which 172 are bots (14.6%).

**Table 1**. Guidelines for the identification of bots

| Process | Description | Evidence |
|---|---|---|
| *Bot label* | Automated account label identification 🤖<br>Example: @MathPHYPapers | |
| *Bot statement* | Bot self-declaration in the account description<br>Example: @mathCObot | |
| *Service-linked accounts* | Auto-generated activity from a service (e.g. repositories)<br>Example: @repositorium | |
| *Activity pattern* | Highly repetitive activity with a clear pattern<br>Example: @ugentbiblio | |
| *Only retweets* | Activity limited to retweeting posts with shared hashtags/words<br>Example: @hecanhazpdf | |
| *Zero interaction* | Limited interaction with the community (replies, likes...)<br>Example: @deep_rl | |
| *High activity* | Very high daily activity<br>Example: @ucl_discovery | |
| *Parallel accounts* | Accounts with a highly similar profile and activity<br>Examples: @Alishba61078926 and @RidaAsghar16 | |
| *Identification* | Profile with limited information (photo, description...)<br>Example: @Saadiii1234 | |
| *Confirmed profile* | Profile with peck check (previous to Twitter Blue)<br>Example: @PeckhamPlatform | |
| *Interactive account* | High human activity (personal tweets, use of emojis...) | |

■ Bot validated   ■ High evidence   ■ Medium evidence   ■ Low evidence   ■ Human validated

Using the sample, we conducted several tests with the aim of optimizing bot detection using BotometerLite. Beyond just the botscore, we took into account the rate at which accounts mention scientific publications, differentiating between total mentions, tweets, and retweets. The intention behind this was to minimize false positives and accurately pinpoint bots that are highly active in the realm of science dissemination. Regarding the botscore, we decided to test three values (0.5, 0.6, and 0.7). This decision was informed by the frequent citation of 0.5 as the standard value for bot identification (Shao et al., 2018; Vosoughi et al., 2018) and the observation that these three values are close to the upper limits of the botscore distribution (Q3[1] + 1.5*IQR[2]) across various ESI fields (Figure 1). Similarly, to establish thresholds for mentions, tweets, and retweets, we assessed the general distribution of such activity at the account level for extreme upper values, multiplying the interquartile range (IQR) by 1.5 (mild outlier) and 3 (extreme outlier).

**Figure 1**. Botscore distribution of tweeters mentioning papers by ESI field

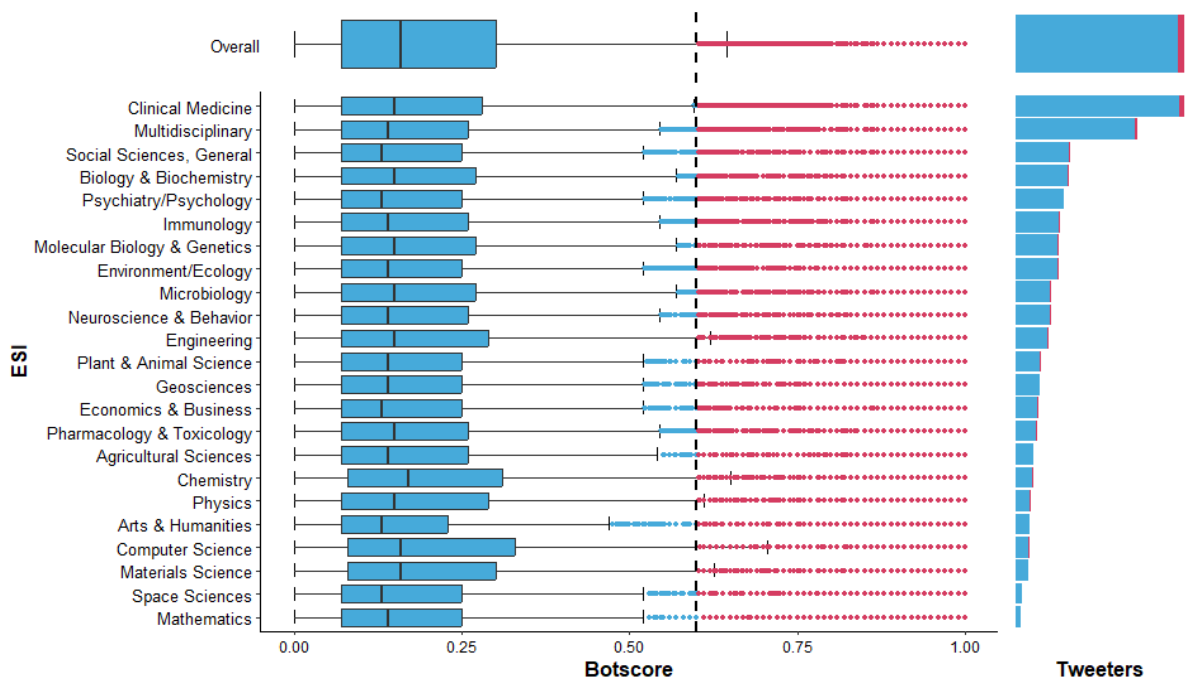

For the optimal selection of parameters, we used several evaluation metrics for classification models. Precision measures the accuracy of positive predictions; recall captures how many true positives were correctly identified; the F1 score harmonizes precision and recall to gauge overall model efficiency; and accuracy evaluates the proportion of total predictions that are correct. As shown in Table 2, the best outcomes, based on F1 scores, are achieved with botscore thresholds of 0.5 and 0.6. Specifically, a botscore greater than 0.5 combined with more than 4 tweets yields an F1 of 0.8690, with precision at 0.7907, recall at 0.9645, and an accuracy of 96.525%—though this accuracy should be considered cautiously due to the correct identification of non-bots, which are the majority. However, we have chosen to use a botscore above 0.6 combined with more than 8 mentions for two main reasons: 1) Cross-referencing our general mentions dataset with those of scholars (Mongeon et al., 2023)[3] and journals (Nishikawa-Pacher, 2023) revealed a significant drop in false positives at a 0.6 threshold compared to 0.5 (Table 3); 2) Filtering tweets does not capture retweet activity, making total mentions more appropriate. Notably, the performance disparity between using a botscore over 0.5 and more than 4 tweets (F1=0.869, accuracy=96.525%) and a botscore of 0.6 with over 8 mentions (F1=0.8294, accuracy=95.678%) is minimal.

---

[1] Third or upper quartile

[2] Interquartile range (IQR) = Q3 − Q1

[3] Only tweeters manually verified by the authors

**Table 2**. Comparative results of precision, recall, F1 and accuracy in the detection of Twitter bots

| Botscore | Activity | Precision | Recall | F1 | Accuracy |
|----------|----------|-----------|--------|--------|----------|
| *0.5* | — | 0.8488 | 0.6697 | 0.7487 | 91.695% |
| *0.5* | *> 8 mentions* | 0.7965 | 0.9514 | 0.8671 | 96.441% |
| *0.5* | *> 13 mentions* | 0.7500 | 0.9627 | 0.8431 | 95.932% |
| *0.5* | *> 2 tweets* | 0.7965 | 0.9514 | 0.8671 | 96.441% |
| *0.5* | *> 4 tweets* | 0.7907 | 0.9645 | 0.8690 | 96.525% |
| *0.5* | *> 6 retweets* | 0.1047 | 0.8182 | 0.1856 | 86.610% |
| *0.5* | *> 9 retweets* | 0.0988 | 0.8500 | 0.1771 | 86.610% |
| *0.6* | — | 0.7733 | 0.7870 | 0.7801 | 93.644% |
| *0.6* | *> 8 mentions* | 0.7209 | 0.9764 | 0.8294 | 95.678% |
| *0.6* | *> 13 mentions* | 0.6744 | 0.9831 | 0.8000 | 95.085% |
| *0.6* | *> 2 tweets* | 0.7209 | 0.9612 | 0.8239 | 95.508% |
| *0.6* | *> 4 tweets* | 0.7151 | 0.9685 | 0.8227 | 95.508% |
| *0.6* | *> 6 retweets* | 0.0988 | 0.9444 | 0.1789 | 86.780% |
| *0.6* | *> 9 retweets* | 0.0988 | 1.0000 | 0.1799 | 86.864% |
| *0.7* | — | 0.5116 | 0.8544 | 0.6400 | 91.610% |
| *0.7* | *> 8 mentions* | 0.4826 | 1.0000 | 0.6510 | 92.458% |
| *0.7* | *> 13 mentions* | 0.4535 | 1.0000 | 0.6240 | 92.034% |
| *0.7* | *> 2 tweets* | 0.4767 | 0.9880 | 0.6431 | 92.288% |
| *0.7* | *> 4 tweets* | 0.4767 | 0.9880 | 0.6431 | 92.288% |
| *0.7* | *> 6 retweets* | 0.0698 | 1.0000 | 0.1304 | 86.441% |
| *0.7* | *> 9 retweets* | 0.0698 | 1.0000 | 0.1304 | 86.441% |

**Table 3**. False positive rate in the detection of Twitter bots using researchers and journals

| Botscore | Activity | Researchers | Journals |
|----------|----------|-------------|----------|
| *0.5* | — | 6.750% | 17.999% |
| *0.5* | *> 8 mentions* | 0.747% | 12.881% |
| *0.5* | *> 13 mentions* | 0.451% | 11.743% |
| *0.5* | *> 2 tweets* | 1.666% | 14.286% |
| *0.5* | *> 4 tweets* | 0.865% | 13.249% |
| *0.5* | *> 6 retweets* | 0.419% | 5.453% |
| *0.5* | *> 9 retweets* | 0.279% | 4.650% |
| *0.6* | — | 3.439% | 9.401% |
| *0.6* | *> 8 mentions* | 0.301% | 6.156% |
| *0.6* | *> 13 mentions* | 0.179% | 5.453% |
| *0.6* | *> 2 tweets* | 0.787% | 6.959% |
| *0.6* | *> 4 tweets* | 0.432% | 6.424% |
| *0.6* | *> 6 retweets* | 0.132% | 1.974% |
| *0.6* | *> 9 retweets* | 0.092% | 1.606% |
| *0.7* | — | 1.628% | 3.948% |
| *0.7* | *> 8 mentions* | 0.127% | 2.108% |
| *0.7* | *> 13 mentions* | 0.077% | 1.874% |
| *0.7* | *> 2 tweets* | 0.391% | 2.543% |
| *0.7* | *> 4 tweets* | 0.216% | 2.375% |
| *0.7* | *> 6 retweets* | 0.042% | 0.502% |
| *0.7* | *> 9 retweets* | 0.032% | 0.368% |

Lastly, to provide additional validation for this criterion, a category from Web of Science was selected to investigate the categorization of accounts as bots. For this exploration, the Mathematics category was chosen, as 275 accounts within it were labeled as bots, accounting for 50% of the mentions and 70% of the tweets. A total of 5 accounts could not be assessed because they had been deleted or set to private by the time of the analysis. Upon manual evaluation of the accounts classified as bots between June 5 and 7, 2023, 219 were indeed found to be automated accounts (81%), while 51 were not. This indicates a high accuracy rate, with the methodology correctly identifying bots responsible for 20,864 tweets/mentions (99.22% of the tweets/mentions attributed to bots). Therefore, we can validate this model and the results derived from its application.

**Table A1**. Top 20 Twitter accounts classified as bots with the most mentions

| Name *Screen name* | MENTIONS | | | | PROFILE | | | |
|---|---|---|---|---|---|---|---|---|
| | **Papers** | **Ment.** | **Tweets** | **RT** | **Foll.** | **Frien.** | **Favs.** | **Status** |
| **Behav Ecol Papers** *@BehavEcolPapers* | 32,678 | 43,538 | 33,637 | 9901 | 4743 | 0 | 0 | 84,668 |
| **Symbiosis papers** *@Symbiosispapers* | 31,653 | 32,792 | 32,792 | 0 | 1254 | 0 | 1 | 72,685 |
| **POPapers** *@geomatlab* | 21,466 | 31,088 | 30,912 | 176 | 1075 | 59 | 8 | 60,405 |
| **UCL Discovery** *@ucl_discovery* | 29,828 | 30,313 | 30,313 | 0 | 2137 | 0 | 15 | 98,547 |
| **agepapers** *@agepapers* | 209 | 30,011 | 30,011 | 0 | 60 | 2 | 0 | 49,663 |
| **Condensed Matter** *@CondensedPapers* | 28,303 | 28,570 | 28,570 | 0 | 3410 | 1 | 0 | 109,951 |
| **African Synchrotron** *@AfSynchrotron* | 27,747 | 28,547 | 27,500 | 1047 | 2989 | 4986 | 56 | 48,349 |
| **Transcriptomes** *@transcriptomes* | 26,027 | 26,717 | 26,717 | 0 | 1866 | 5 | 1 | 52,707 |
| **RRID Robot** *@RobotRrid* | 18,077 | 26,324 | 26,324 | 0 | 140 | 0 | 0 | 36,034 |
| **neopapers** *@neo_papers* | 3150 | 22,745 | 22,740 | 5 | 3568 | 0 | 4 | 74,094 |
| **SystematicReviewBot** *@EvidenceRobot* | 8927 | 20,111 | 0 | 20,111 | 7412 | 1 | 58 | 266,449 |
| **arXiv astro-ph** *@_arXiv_astro_ph* | 17,026 | 19,498 | 19,498 | 0 | 1043 | 12 | 1 | 48,199 |
| **Microbiome Articles** *@microbe_article* | 17,898 | 18,567 | 18,565 | 2 | 5848 | 1 | 1 | 35,807 |
| **pseudomonaspapers** *@pseudo_papers* | 17,437 | 17,533 | 17,533 | 0 | 1335 | 34 | 0 | 38,889 |
| **Lymphoma Papers** *@Lymphoma_Papers* | 13,956 | 16,044 | 16,029 | 15 | 2191 | 0 | 19 | 40,379 |
| **ImmunologyPapers** *@Immunol_papers* | 13,972 | 15,350 | 15,350 | 0 | 1221 | 0 | 0 | 77,839 |
| **AtmosSciBot** *@AtmosSciBot* | 11,716 | 14,788 | 14,788 | 0 | 414 | 1 | 2 | 23,526 |
| **Mycobacterium Papers** *@MycobactPapers* | 10,047 | 14,389 | 14,385 | 4 | 1492 | 6 | 7 | 25,503 |
| **painmanwise** *@painmanwise* | 10,453 | 13,682 | 13,682 | 0 | 225 | 15 | 0 | 51,264 |
| **arXiv gr-qc** *@arXiv_grqc* | 9224 | 13,637 | 13,637 | 0 | 221 | 0 | 0 | 37,785 |

**Tabla S2**. Difference between bots and non-bots in the percentage distribution of papers mentioned by type of open access

| Non-bot ▭ Bot | **Gold** | **Green** | **Hybrid** | **Bronze** | **Closed** |
|---|---|---|---|---|---|
| *Agricultural Sciences* | -3.60% | 1.40% | -0.10% | -0.50% | 4.50% |
| *Arts & Humanities* | -1.00% | 12.70% | 4.20% | 0.50% | -10.50% |
| *Biology & Biochemistry* | -1.70% | -2.60% | -0.80% | -0.20% | 2.80% |
| *Chemistry* | 0.70% | 4.50% | -1.10% | -0.30% | -1.70% |
| *Clinical Medicine* | 1.30% | 2.20% | 0.20% | -0.10% | -1.90% |
| *Computer Science* | -0.30% | 20.90% | 0.80% | 1.60% | -15.00% |
| *Economics & Business* | 1.50% | 10.70% | 2.90% | 3.50% | -11.40% |
| *Engineering* | -9.30% | 5.20% | -1.20% | -0.30% | 1.50% |
| *Environment/Ecology* | 2.30% | 10.20% | 0.60% | 1.20% | -4.30% |
| *Geosciences* | 9.20% | 14.90% | 1.20% | 2.60% | -13.50% |
| *Immunology* | -4.20% | -6.00% | -1.20% | -0.70% | 6.40% |
| *Materials Science* | -12.70% | 5.20% | 0.20% | -1.20% | 3.30% |
| *Mathematics* | -6.20% | 28.40% | -4.90% | 2.80% | -20.60% |
| *Microbiology* | -3.50% | -4.80% | -0.90% | -0.30% | 4.70% |
| *Molecular Biology & Genetics* | -1.80% | -3.60% | -1.60% | 0.10% | 3.30% |
| *Multidisciplinary* | -3.10% | 0.70% | 0.80% | 0.70% | 0.00% |
| *Neuroscience & Behavior* | -5.70% | -2.00% | 0.80% | -0.20% | 3.00% |
| *Pharmacology & Toxicology* | -2.60% | -5.00% | -1.30% | 0.00% | 4.60% |
| *Physics* | -5.10% | 24.20% | 3.20% | -0.90% | -16.90% |
| *Plant & Animal Science* | 1.40% | 4.70% | -0.10% | 0.10% | -1.40% |
| *Psychiatry/Psychology* | -1.00% | 2.80% | 2.10% | 0.40% | -2.50% |
| *Social Sciences. General* | 2.20% | 11.60% | 2.30% | 1.40% | -10.70% |
| *Space Sciences* | -0.60% | 5.10% | 5.30% | -4.40% | -3.50% |